# RESEARCH OF INFLUENCE OF TEMPERATURE DEFORMATIONS OF THE BIG ELASTIC ELEMENTS ON DYNAMICS OF A SPACE VEHICLE

**Sedelnikov A.V., Kazarina M.I.**

**Samara state aerospace university**

Important characteristic is the module of micro-acceleration inside the operating area of technological equipment of the modern space laboratory [1]. Studies show that perturbations of a domestic disturbance make the largest contribution to the field of micro-accelerations. Most significant among them is the work of the governing rocket motor of attitude and motion control system of the spacecraft. It is necessary to limit accelerations for the success of the technological processes. This can be achieved by different design methods.

Reducing of contribution to the overall level of micro-accelerations perturbations from domestic disturbance raises a number of other urgent tasks. It is required to consider other disturbing factors to create a reliable assessment.

SkyLab is periodically exposed to high and low temperatures during its orbital motion. Large elastic elements are influenced by significantly uneven temperature field. Subsequently, large elastic elements abruptly change their shape, size and physical properties. This phenomenon corresponds to the specific percussion effects, which is transmitted to the spacecraft. Percussion effects leads to additional micro-accelerations. I appreciate their value in my research.

The maximum allowable level of micro-accelerations amounts up to 10 m / s$^2$ for the modern space laboratory, such as "OKA-T". Therefore, the task of research and evaluation of micro-accelerations caused by temperature fluctuations elastic elements of the spacecraft is of great urgency.

As a first approximation in the calculation it is estimated micro acceleration of a rod of the elastic element. Such a model can match different kinds of antennas.

Consider the heat equation [2]:

$$\kappa\nabla T + B = c\dot{T}, \qquad (1)$$

where $\kappa$ - thermal conductivity; B - the amount of heat; c - heat capacity per unit of volume. Boundary conditions are the following:

$$T / O_1 = T_1, \quad \kappa\partial_n T / O_2 = q.$$

Here, the temperature and the external heat flux are set.

The disadvantage of (1) is non-wave nature and infinitely large velocity of spread of heat. Wave type equations are obtained by means of complicating of heat flux vector $\boldsymbol{h}$:

$$\Theta\boldsymbol{h} + \boldsymbol{h} = -\kappa\nabla T. \qquad (2)$$

Here $\Theta$ - time constant of establishing of heat flow; $\boldsymbol{h}$ reacts to change $\nabla T$ with delay. Then the equation:

$$-div\,\boldsymbol{h} + B = c\dot{T}$$

with the ratio (2) leads to the hyperbolic equation:

$$\kappa\nabla T + B + \Theta\dot{B} = c(\dot{T} + \Theta\ddot{T}).$$

Rate of heat distribution is equal to $\sqrt{\kappa / c\Theta}$.

To obtain an approximate estimate of the model we use a homogeneous rod model of length $l$ and insulated side surface. The initial temperature distribution is selected in accordance with the conditions of thermal phenomenon. It takes into account real features of heating of large elastic elements of the spacecraft. The boundary conditions are variable in nature. When heating the temperature of the ends of the rod will vary:

$$\frac{\partial T}{dt} = a^2 \frac{\partial^2 T}{\partial x^2}, \qquad (3)$$

where $a=\sqrt{\dfrac{\lambda}{c\rho}}$, $\rho$ - density, c - specific heat, $\lambda$ - thermal conductivity.

In case of the problem to be solved boundary conditions and initial temperature distribution can be written as follows:

$$T(o,t)=\mu_1(t), T\left(l,t\right)=\mu_2(t), \quad 0<t<+\infty,$$

$$T(x,0)=\varphi(x), \quad 0<x<l,$$

where $\varphi(x)$ - function corresponding to the conditions of heat stroke. It was chosen in line with the face-centered cubic lattice structure of aluminum with a parameter 4,050 · 10-4 m.

According to the Fourier method, the solution (3) takes the following form:

$$T(x,t)=u(x,t)+v(x,t),$$

where $u(x,t)=\mu_1(t)+\dfrac{x}{l}\left[\mu_2(t)-\mu_1(t)\right]$, $v(x,t)=\sum\limits_{n=1}^{\infty} C_n e^{-\left(\frac{\pi n}{l}\right)^2 a^2 t}\sin\dfrac{\pi n}{l}x$,

$$C_n=\frac{2}{l}\int\limits_0^l \varphi\left(\xi\right)\sin\frac{\pi n}{l}\xi d\xi .$$

We make two estimates of heating time for the case of a regular mode of large elastic elements. Let's regard solar panel as a real constructive element. Cosine of the angle between the normal to the SP and the direction of the Sun must be at least 0,9. The second estimate is for a maximum duration of heating of large elastic elements. Let's consider a real constructive element of radiator panel.

Calculate the incident heat flux, referable to the elastic element by means of formula evaluation:

$$\Phi_n=\frac{\Phi}{4\pi}\Omega,$$

where $\Phi$ - full flow of energy from the Sun, $\Omega=\dfrac{S}{R^2}\cos\alpha$ - the solid angle, $R$ - the distance from the Sun to the Earth.

Numerical simulation of full-time heating in the package of Mathcad gave us the results shown in Fig. 1 and 2.

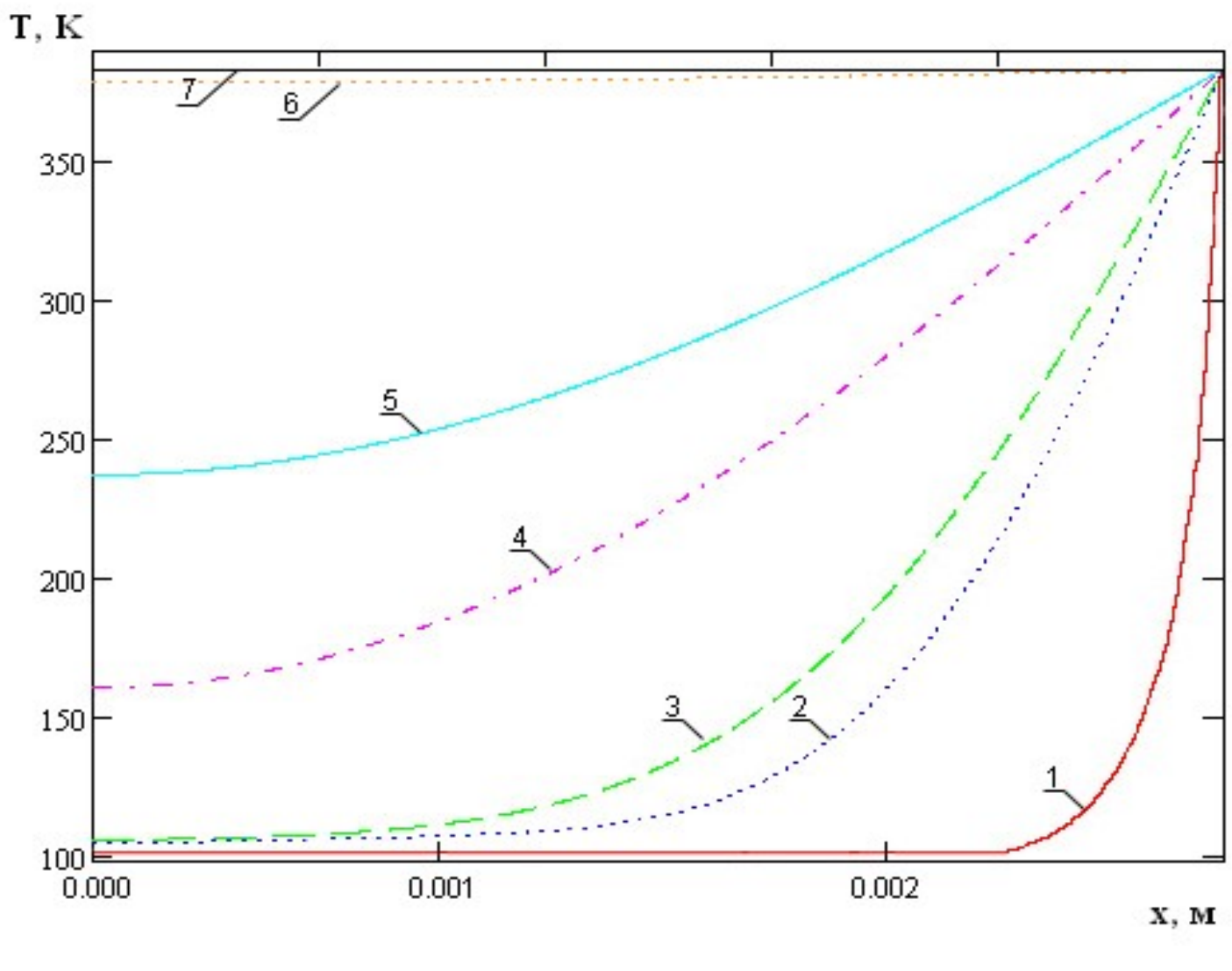

Fig. 1. - Variation of temperature distribution inside the SP in case of regular work

In Fig. 1 curve 1 represents the initial distribution of temperatures $\varphi(x)$, as described above. Next diagram is aligned with the passage of time and becomes a horizontal line 7, indicating uniform heating of the SP. Lines 1, 2, 3, 4, 5, 6 and 7 correspond to time points 0, with 0,005; 0,01 s, 0,05 s, 0,1 s, 0,5 s and 1,6 s respectively. Thus, one can assert that the evaluation of the full-time warm SP is as follows:

$$\Delta t = 1{,}6c \tag{4}$$

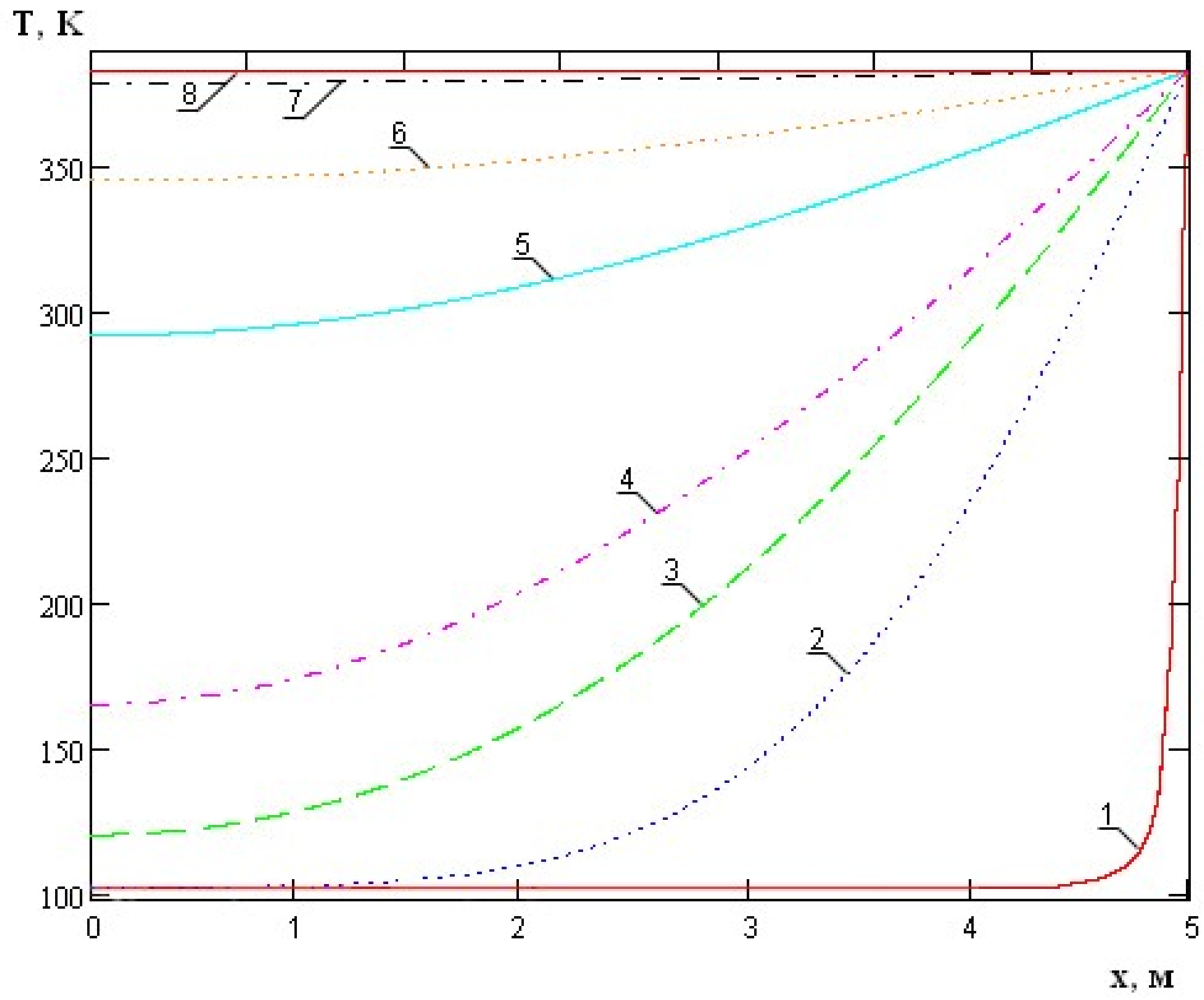

Fig. 2. - Variation of temperature distribution inside the radiator

In Fig. 2 it is similar to Fig. 1 curve 1 represents the initial distribution of temperatures $\varphi(x)$, and lines 1, 2, 3, 4, 5, 6, 7 and 8 correspond to time points 0, 2 h 46 m 40 s (104 s), 13 h 53 m 20 s (5 · 104 c), 27 h 46 m 40 s (105) 3 days 9 h 20 m 00 with (3 · 105) 5 days 18 h 53 m 20 s (5 · 105) 11 days 13 hrs 46 meters 40 (106) and 35 d 9 h 18 m 52 sec. Moreover, the last one is a full time of heating seven-meter radiator through its butt.

The heating is insignificant. In most practical problems it can be neglected. This time estimate is overstated because of non-counting of emission of heat elastic element. This estimate was obtained by heating due to heat passing through the end section of the element.

Let's estimate accelerations, which could result from the temperature deformations of the SP for the space lab type "Nika-T".

We estimate the momentum in the case of a linear extension:

$$dp = \mu V_0\left(2 - \frac{x}{l}\right)dx, \quad p = \mu V_0\left(2x - \frac{x^2}{2l}\right)\Bigg|_0^{\frac{l}{2}} = \frac{3}{8}\mu l V_0,$$

where $\mu$ - mass per unit of length.

Increment of pulse SP due to its deformation is approximately equal to 2.2 kg m / sec. Assuming that all the momentum will be transmitted to the spacecraft body without loss one can estimate the module of created micro accelerations:

$$w = \frac{\Delta p}{m_0 \cdot \Delta t} \approx 2 \cdot 10^{-4} \ м/c^2 \tag{5}$$

This estimate (5) is somewhat overvalued. However, it should be noted that the requirements for the accelerations in the "Nick-T" are no more $2 \cdot 10^{-5}\ m/\sec^2$, that is an order of magnitude below the estimate (5).

The significance of the influence of thermal strains allowed us to continue the study as a part of complex three-dimensional model. It was remarkable that three-dimensional model of a homogeneous orthotropic plate was used instead of the rod model. Program LS-DYNA carried out numerical calculations for the layout of space laboratory-type "Nika-T". These results are comparable with the calculations on a simpler layout.

Our studies suggest the relevance of the problem of the influence of temperature on the deformation dynamics of spacecraft motion for specialized spacecraft technological goals. Without such assessments we cannot state that there are conditions of microgravity calm during the process, if the spacecraft periodically appears in the shadow and out of it.

Address for connection: Russia, state Samara, 443026, p.b. 1253, Dr. Andry Sedelnikov
Address on Russian: Россия, г. Самара, а/я 1253.
E-mail: axe_backdraft@inbox.ru